\documentclass[preprint,pre,longbibliography]{revtex4-1} 

\usepackage{natbib}
\usepackage{graphicx}
\usepackage{amssymb}
\usepackage{amsfonts}
\usepackage{amsmath}
\usepackage{amsbsy}
\usepackage{mathrsfs} 
\usepackage{color}
\usepackage{gensymb} 
\usepackage{hyperref} 
\usepackage{soul}%\st{...}
\usepackage{esint}
\usepackage{mymacros}
\def\XXint#1#2#3{{\setbox0=\hbox{$#1{#2#3}{\int}$}
     \vcenter{\hbox{$#2#3$}}\kern-.5\wd0}}

\begin{document} 

\title{Absorption characteristics of large acoustic metasurfaces} 

\author{Ory Schnitzer}
\author{Rodolfo Brand\~{a}o}
\affiliation{Department of Mathematics, Imperial College London, 180 Queen's Gate, London SW7 2AZ, UK}

\begin{abstract}
Metasurfaces formed of arrays of subwavelength resonators are often tuned to `critically couple’ with incident radiation, so that at resonance dissipative and radiative damping are balanced and absorption is maximised. Such design criteria are typically derived assuming an infinite metasurface, whereas the absorption characteristics of finite metasurfaces, even very large ones, can be markedly different in certain frequency intervals. This is due to the excitation of surface waves, intrinsic to resonant metasurfaces, and especially meta-resonances, namely collective resonances where the surface waves form standing-wave patterns over the planar metasurface domain. We illustrate this issue using a detailed model of a Helmholtz-type acoustic metasurface formed of cavity-neck pairs embedded into a rigid substrate, with geometric and dissipation effects included from first principles (R.~Brand{\~a}o and O.~Schnitzer, Wave Motion, \textbf{97} 102583, 2020). \end{abstract}

\maketitle

\section{Introduction}
Designing surfaces to effectively absorb wave energy is a classical engineering problem, with many applications across wave physics including room acoustics, noise suppression, wireless power transfer, targeted heating, sensing and filtering. Naturally, over the last two decades this endeavour has been significantly influenced by developments in the field of metamaterials, where novel paradigms for wave control are realised in practice with engineered micro-structured bulk media and surfaces incorporating subwavelength resonators. Examples include thin absorbing metasurfaces \cite{Li:16,Zhang:16,Chen:17}; critically coupled metasurfaces for perfect absorption \cite{Romero:16,Romero16b,Jimenez:16} and generalisations based on coherent-perfect absorption \cite{Wei:14} and their exceptional points \cite{achilleos:2017,Wang:21}; graded-metasurface `rainbow' absorbers \cite{Jimenez:17}; and complex unit-cell designs incorporating multiple resonators for near-perfect broadband absorption \cite{Li:16b,Jimenez:17b,Peng:18}. 

Metasurface absorbers are typically designed based on theoretical analyses of infinite surfaces, or, in the other extreme, a small number of resonators in a confined geometry. Accordingly, experimental demonstrations often employ resonator-loaded channels, surfaces decorated with strongly lossy edges, or small surface samples spanning the entire cross section of an impedance tube \cite{Kim:06,Li:16b,Romero16b}. The latter experimental technique mimics the theoretical scenario of a plane wave normally incident on an infinite surface. `Real-world'  scenarios, however, involve a surface covered by one or more absorbing patches, in which case the finiteness of the patches is likely important. Indeed, diffractive edge effects have long been known to be important for the characterisation of conventional acoustic absorbers, in particular for resolving the discrepancy between measured and theoretical absorption coefficients \cite{Thomasson:80,Thomasson:82,Brandao:12}. 

We shall argue that finite-size effects can be especially pronounced in the case of metasurface absorbers formed of arrays of subwavelength resonators, especially in regions of parameter space which are traditionally ignored in the absorption context. Thus, it is well known that infinite arrays of subwavelength resonators almost always support surface waves in certain frequency intervals. For infinite surfaces, these can be asymptotically excited by incident beams of finite width, or localised sources; for finite surfaces, these can also be asymptotically excited by incident plane waves via edge diffraction. Furthermore, for sufficiently small, or low-loss, metasurfaces the surface waves can have such large propagation lengths that they reflect at the edges of the metasurface (as well as refract into the bulk) multiple times before significantly attenuating. The surface waves can then form standing waves, thence giving rise to collective resonances of the metasurface as a whole that we shall call `meta-resonances'. The effects of meta-resonances on scattering have been observed in acoustics using a metasurface formed of arrays of soda cans acting as low-loss Helmholtz resonators \cite{Lemoult:16} and in photonics using arrays of nanoplasmonic resonators \cite{Yang:20}, however the focus in these experiments is on  scattering rather than absorption. They have also been described theoretically \cite{Thompson:08,Schnitzer:19a}, albeit in the absence of loss. 

Metasurface absorbers are usually designed to `critically couple' with incident radiation. This means tuning material loss to balance dissipation and radiation damping at resonance. Roughly speaking, material loss should be low enough to allow for strong resonances, but not too low so that energy is mainly lost via dissipation rather than by energy leakage to infinity. Critical coupling conditions for metasurface absorbers, however, are typically derived in the idealised scenario of plane waves incident on an infinite metasurface, in which case surface waves are not excited. With regards to finite metasurface absorbers, this observation gives rise to important questions. Thus, for finite metasurfaces with material loss tuned based on critical-coupling theory for infinite metasurfaces, what is the effect of surface-wave excitation as a function of metasurface size? Second, do finite low-loss metasurfaces possess, against conventional wisdom, useful absorption characteristics associated with surface-wave excitation?

These questions appear to have received little consideration in the literature. Perhaps this is because direct numerical simulations of large metasurfaces are very expensive, a main challenge being the need to resolve multiple disparate length scales. Thus, in a typical scenario, the linear dimension of the metasurface is either large or comparable to the free-space wavelength, which is much larger than the subwavelength scales of the resonators and the spacing between the resonators. Moreover, the resonators themselves often involve complex and multiple-scale geometries, as in the case of Helmholtz or space-coiled resonators in acoustics, split-ring resonators and near-singular plasmonic nanostructures in photonics. Lastly, dissipation effects often involve even smaller length scales, for example the width of viscous and thermal boundary layers in acoustics or skin depth in photonics. 
While effective-media, or homogenised, models offer immense simplification, these generally fail in the case of low-loss metasurfaces owing to the excitation of deeply subwavelength surface waves.

The aim of this paper is to theoretically demonstrate the substantial and perhaps unexpectedly singular effect of surface-wave excitation on the absorption characteristics of finite, even very large, metasurface absorbers. To this end, we shall use a reduced multiple-scattering model of a finite-sized acoustic Helmholtz-type metasurface formed of cavity-neck pairs which are embedded into a flat rigid substrate. We have recently derived this model using matched asymptotic expansions \cite{Brandao:20}, first considering the acoustic response of a single cavity-neck resonator and then generalising to an arbitrary planar distribution of resonators based on a Foldy-type multiple-scattering  approximation \cite{Foldy:45,Martin:06}. The geometric details of the three-dimensional cavities and necks, which can be quite general, are encoded in lumped parameters, which are defined systematically in terms of boundary-value problems and for which we have obtained numerical values and analytical approximations \cite{Brandao:20,Brandao:20Imp}. The model also allows for an arbitrary geometric arrangement of non-identical resonators, though here we will focus on a square lattice of identical resonators. Most importantly, and in contrast to preceding models of Helmholtz resonators, the model derived in \cite{Brandao:20} includes dissipative effects from first principles; these are shown to be dominated by the viscous boundary layers in the necks of the resonators. While Helmholtz-type metasurfaces are relatively basic compared to more sophisticated concepts \cite{Yang:17,Assouar:18}, they remain very common in applications: essentially similar surfaces are used as absorbing panels for indoor acoustics and as acoustic liners for jet engines; the soda-can metasurface \cite{Lemoult:16} already mentioned provides a particularly simple realisation. 

The paper is structured as follows. In \S\ref{sec:formulation}, we formulate a multiple-scattering model of the Helmholtz-type acoustic metasurface based on the theory developed in \cite{Brandao:20,Brandao:20Imp}, as well as derive a homogenisation approximation of that model which is valid in some cases. In \S\ref{sec:cc}, we derive explicit critical-coupling conditions in the two extreme cases of a single resonator and an infinite metasurface. In \S\ref{sec:sw}, we study the surface waves supported by an infinite metasurface and their attenuation.  In \S\ref{sec:large}, we numerically explore the absorption characteristics of large metasurfaces, comparing both `tuned' (critically coupled) and `detuned' (low-loss) metasurfaces to their infinite counterparts and attempt to qualitatively explain some of the observed characteristics. In \S\ref{sec:conc}, we give concluding remarks and propose directions for further study. 

\section{Helmholtz-type metasurface}\label{sec:formulation}
\subsection{Geometry and physical assumptions}
We consider a model acoustic metasurface formed of $N$ identical cavities which are embedded into a flat substrate and arranged in a square lattice. Each cavity (volume $l^3$) is connected to the half-space exterior to the substrate by a cylindrical neck (radius $\epsilon l$,  length $2\epsilon hl$) whose axis is perpendicular to the substrate plane. The shape of the cavity is arbitrary, except that its boundary is assumed flat in the close vicinity of the neck, which does not protrude into the cavity or exterior half-space. The spacing between the necks is denoted $A^{1/2}l$, such that $Al^2$ is the area of a unit cell of the lattice. The fluid filling the cavities, neck and the exterior half-space is assumed to be a viscous and thermally conducting gas (speed of sound $c$, density $\rho$, kinematic viscosity $\nu$). The substrate is assumed rigid, isothermal, with the fluid velocity satisfying a no-slip boundary condition. Accordingly, there is an additional characteristic length scale, $l_v=\sqrt{\nu/\omega_H}$, corresponding to the width of the viscous boundary layer, $\omega_H$ being a  characteristic angular frequency to be specified below; the characteristic width of the thermal boundary layer is assumed comparable, as is the case for air. Associated with $l_v$ is the ratio $\delta=l_v/l$, which serves as a measure of material loss.     

\begin{figure}[t!]
\begin{center}
\includegraphics[trim=30 0 20 20,scale=0.45]{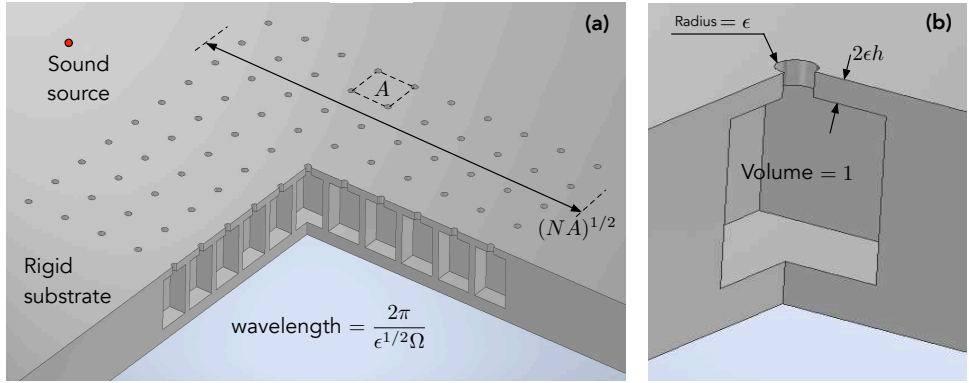}
\caption{(a) Dimensionless sketch of the Helmholtz-type acoustic metasurface  in the case of cubic cavities, with lengths  normalised such that the cavities have unit volume. (b) Geometry of the cavity-neck resonators.}
\label{fig:sketch}
\end{center}
\end{figure}

We adopt a dimensionless convention where lengths are normalised by $l$. The geometry is shown in Fig.~\ref{fig:sketch} for the case of cubic cavities. Relative to an arbitrary fixed origin, we denote the dimensionless position vector by $\br$ and position vectors of the neck centres in the substrate plane by $\br_n$, where $n=1,2,\ldots,N$. Furthermore, we denote the vertical coordinate measured from the substrate plane by $z$. The pressure field will be sought as the real part of $p_{\infty}p(\br)\exp(-i\omega t)$, where $p_{\infty}$ is a reference magnitude, $\omega$ is the angular frequency, $t$ is time and the reduced pressure field $p(\br)$ is a dimensionless and complex-valued phasor field.

Our interest is in studying the sound-absorption characteristics of the metasurface defined above for $\epsilon\ll 1$. In that limit, each cavity-neck pair is expected to behave like a Helmholtz resonator of resonance frequency on the order of $\omega_H=\epsilon^{1/2}c/l$, implying wavelengths of order $\epsilon^{-1/2}$ relative to the linear cavity dimension $l$. This scaling motivates the definition of the dimensionless frequency $\Omega=\omega/\omega_H$. 

\subsection{Multiple-scattering model}\label{ssec:foldy}
In \cite{Brandao:20}, we first studied the  limit $\epsilon\ll1$ in the case of a single cavity-neck resonator, with $\Omega\simeq 1$ and $\delta\ll\epsilon$, the symbol $\simeq$ henceforth standing for `of asymptotic order.' The condition $\delta\ll\epsilon$, namely that the viscous and thermal boundary layers are thin relative to the neck radius, was shown to be equivalent to saying that dissipation is sufficiently weak such that the cavity-neck pair exhibits asymptotically significant resonance. Furthermore, dissipation is then only important in a vicinity of the resonance frequency, with viscous effects dominant over thermal effects and contributed mainly by the viscous boundary layers close to the neck. By analysing a series of distinguished limits in the $\delta$--$\Omega$ parameter space, representing different levels of loss and frequencies increasingly close to resonance, we systematically derived a `unified' asymptotic model for the acoustic response of a single cavity-neck resonator that holds to leading asymptotic order throughout the regime $\Omega\simeq 1$ and $\delta\ll\epsilon$. We then presented an intuitive generalisation of this unified model to an arbitrary number and arrangement of resonators, based on a Foldy-type multiple-scattering approximation  \cite{Foldy:45,Martin:06} and assuming $A^{1/2}\gg \epsilon$, namely that the separation between the necks is large relative to their radius. This includes the case $A\simeq 1$, where the separation is subwavelength and comparable in size to the cavity. 

In what follows, we formulate a multiple-scattering model based on the theory developed in \cite{Brandao:20,Brandao:20Imp}. As $\epsilon\to0$, the neck openings in the substrate plane shrink to the points $\br_n$ and the reduced pressure $p$ satisfies the reduced-wave equation
\begin{equation}\label{Helm p}
\nabla^2p + \epsilon\Omega^2p=f(\br) \quad \text{for} \quad z>0,
\end{equation}
where $f(\br)$ accounts for a possible sound sources in the exterior half-space. On the substrate plane, $p$ satisfies the Neumann boundary condition 
\begin{equation}
\pd{p}{z}=0 \quad \text{at} \quad z=0 \quad (\br\ne\br_n \quad \text{for} \quad n=1,2,\ldots,N),
\end{equation}
whereas, as $\br\to\br_n$ in the exterior half-space, $p$ approaches monopolar singularities whose amplitudes satisfy point constraints specified below. The scattered field, $p(\br)-p\ub{i}(\br)$, satisfies an outward radiation condition as $z\to\infty$, wherein the incident field $p\ub{i}(\br)$ is the reduced pressure in the absence of the substrate, associated with incoming far-field radiation and the  source distribution $f(\br)$. 

In light of the above, the reduced pressure is approximated as
\begin{equation}
p=p\ub{i}(\br)+p\ub{r}(\br)+\sum_{n=1}^N \alpha_n\frac{e^{i\epsilon^{1/2}\Omega|\br-\br_n|}}{|\br-\br_n|},
\end{equation}
where $\alpha_n$ are diffraction coefficients associated with  spherical waves emitted from the necks and we introduce the reflected field $p\ub{r}(\br)$ defined as the solution for the reduced pressure in the absence of the cavity-neck resonators. The reflected field can readily be found for an arbitrary incident field using the method of images. We shall refer to the sum $p\ub{a}(\br)=p\ub{i}(\br)+p\ub{r}(\br)$ as the ambient field. In our examples, we consider two cases: (i) normally incident plane wave, in which case $p\ub{i}(\br)=e^{-i\epsilon^{1/2}\Omega z}$ and $p\ub{a}(\br)=2\cos(\epsilon^{1/2}\Omega z)$; and (ii) a monopole source located at $\br=\br_s$, in which case $p\ub{i}$ is the free-space Green's function $G(\br,\br_s)=|\br-\br_s|^{-1}e^{i\epsilon^{1/2}\Omega|\br-\br_s|}$ and $p\ub{a}(\br)=G(\br;\br_s)+G(\br;\br_s')$, $\br_s'$ being the reflection of $\br_s$ with respect to the substrate plane.

The $n$th diffraction coefficient is determined by requiring that the resonator associated with it reacts as if it was isolated and exposed to the effective ambient pressure field
\begin{equation}
p\ub{a}_n(\br) = p\ub{a}(\br)+\sum_{\substack{m=1\\m\ne n}}^N \alpha_m\frac{e^{i\epsilon^{1/2}\Omega|\br-\br_m|}}{|\br-\br_m|}, \quad n=1,2,\ldots,N.
\end{equation}
i.e., the sum of the ambient pressure and the spherical waves diffracted from all of the other resonators. As already mentioned, the response of a single embedded resonator was asymptotically analysed in \cite{Brandao:20}. Thus, consistently with the standard picture of a Helmholtz resonator, the pressure in each cavity is approximately uniform. Furthermore, the cavity pressures, denoted $p_n$, are proportional to their associated diffraction coefficients according to
\begin{equation}\label{alpha p}
\alpha_n=\frac{\epsilon\Omega^2}{2\pi}p_n, \quad n=1,2,\ldots,N.
\end{equation}
This relation physically represents compression and expansion of the gas in the cavities. In the single-resonator theory, the cavity pressure is found to be proportional to the ambient pressure. Thus, for $N$ resonators, the cavity pressures are calculated by solving the system of equations 
\begin{equation}
g^{-1}p_n=p\ub{a}(\br_n)+\frac{\epsilon\Omega^2}{2\pi}\sum_{\substack{m=1\\m\ne n}}^N p_m\frac{e^{i\epsilon^{1/2}\Omega |\br_m-\br_n|}}{ |\br_m-\br_n|}, \quad n=1,2,\ldots,N,
\end{equation}
where the factor of proportionality  
\begin{equation}\label{g def}
g=-\frac{K}{\Omega^2-\bar{\Omega}^2+i(\Delta_r+\Delta_v)}
\end{equation}
comes from the asymptotic single-resonator theory. Here $\bar{\Omega}$ is the resonance frequency
\begin{equation}\label{bar Omega}
\bar{\Omega}^2=K+\epsilon K^2\sigma-\bar{\Delta_v},
\end{equation}
$\Delta_r$ and $\Delta_v$ are radiative and viscous damping factors defined as  
\refstepcounter{equation}
$$
\label{damping factors}
\Delta_r=\frac{\epsilon^{3/2}\Omega^3K}{2\pi}, \quad \Delta_v=\frac{\delta}{\epsilon}\frac{\Omega^{3/2}K\Theta }{2^{1/2}},
\eqno{(\theequation{\mathrm{a},\!\mathrm{b}})}
$$
respectively, $K,\Theta$ and $\sigma$ are lumped parameters, and $\bar{\Delta}_r$ and $\bar{\Delta}_v$ are, respectively, $\Delta_r$ and $\Delta_v$ evaluated at $\Omega=\bar{\Omega}$.

The parameters $K$ and $\Theta$ are functions of the aspect ratio $h$ of the neck. In particular, $K$ is the so-called Rayleigh conductivity of the neck normalised by its radius \cite{Howe:book}, which is determined by solving a potential-flow problem involving the neck geometry. The parameter $\Theta$, which captures the role of the neck geometry on viscous resistance, is calculated as a quadrature of that same potential flow, representing the displacement of that flow by the viscous boundary layer. In \cite{Brandao:20Imp}, we derived accurate asymptotic approximations for $K$ and $\Theta$ in the limits of small and large $h$, corroborating and significantly extending previous approximations in the literature; we also provided numerical results for arbitrary $h$. The parameter $\sigma$ is a cavity shape factor whose calculation entails solving a Poisson-type boundary value problem defined over the cavity domain with the neck opening reduced to a point; in \cite{Brandao:20}, this problem is defined and solved for several families of cavity shapes. Expressions and numerical results for $K,\Theta$ and $\sigma$ are provided in the supplementary material \cite{SM_ar}. 

The quantity of interest for us is the power dissipated by the metasurface, or the averaged dissipation per resonator which can be thought of as an absorption efficiency. We shall normalise power by $p_{\infty}^2l^2A/2\rho c$, the incident power per unit area of the metasurface in the case of a normally incident plane wave. Once the multiple-scattering problem is solved for a given incident field, the total dimensionless dissipation, denoted $\mathcal{D}$, is provided as
\begin{equation}\label{diss def}
\mathcal{D}=\frac{\epsilon^{1/2}\Omega \Delta_v}{KA}\sum_{n=1}^N|p_n|^2.
\end{equation}

Formula \eqref{diss def} is derived in the supplementary material \cite{SM_ar} in two independent ways. First, it is shown that within the framework of the multiple-scattering model \eqref{diss def} gives exactly the rate of energy leakage through the limiting neck positions $\br_n$. Second, it is shown that \eqref{diss def} asymptotically agrees, to leading order both near and away from resonance, with a direct integration of the viscous dissipation in the boundary layers near the necks. The latter calculation is carried out by relating the cavity pressures $p_n$ in the multiple-scattering model to the matched asymptotic expansions developed in  \cite{Brandao:20,Brandao:20Imp} to describe the cavity, neck and exterior regions. 

We note that the multiple-scattering model herein slightly differs from that in \cite{Brandao:20}. The difference is that in \cite{Brandao:20} the frequency dependence of $\Delta_r$ and $\Delta_v$ is ignored, with $\Omega$ in \eqref{damping factors} approximated by the leading-order resonance frequency $K^{1/2}$. This results in a mathematically more explicit formulation and the difference is asymptotically negligible for the purpose of calculating pressure to leading order both near and away from resonance, as well as dissipation near resonance. Nonetheless, we choose to introduce this frequency dependence here as the analysis in the supplementary material \cite{SM_ar} shows that this extends the asymptotic validity of \eqref{diss def} to off-resonance frequencies and ensures that the multiple-scattering theory and \eqref{diss def} are exactly self-consistent both physically and numerically. 

For the numerical examples in this paper we shall assume a subwavelength unit-cell area $A=1$, neck parameters $\epsilon=0.02$ and $h=2$, for which $K\approx 0.56$ and $\Theta\approx 3.71$, and a cubic cavity for which $\sigma\approx 0.2874$. We will vary the number of resonators $N$ and use $\delta$ to tune the material loss.

\subsection{Homogenised model} \label{ssec:homo}
The multiple-scattering model can in some cases be approximated by a homogenised model  where the metasurface is represented by an effective impedance condition. A necessary, but as we shall see not sufficient, condition is subwavelength spacing, i.e., $A^{1/2}\ll1/\epsilon^{1/2}$. Let $P$ be a `macroscale'  exterior pressure, assumed to vary on some long length scale $L\gg A^{1/2}$, which may or may not be comparable to the order $1/\epsilon^{1/2}$ free-space wavelength. On this length scale, the substrate plane $z=0$ decomposes into a metasurface patch $\mathcal{M}$ (a square of area $\mathcal{A}=NA$) and the remaining domain, say $\mathcal{S}$. The homogenised problem consists of the reduced-wave equation
\begin{equation}\label{homo eq}
\nabla^2P+\epsilon\Omega^2P=0 \quad \text{for} \quad z>0,
\end{equation}
the impedance boundary condition
\begin{equation}\label{imp}
\pd{P}{z}+\frac{\epsilon \Omega^2 g}{A}P=0 \quad \text{on} \quad  \mathcal{M},
\end{equation}
the Neumann boundary condition 
\begin{equation}
\pd{P}{z}=0 \quad \text{on} \quad \mathcal{S}
\end{equation}
and an outward radiation condition on the scattered field $P-P\ub{i}$, where $P\ub{i}=p\ub{i}$ from the multiple-scattering model. The averaged dissipation per resonator is calculated as 
\begin{equation}\label{diss hom}
\mathcal{D}_H/N=\frac{\epsilon^{1/2}\Omega \Delta_v|g|^2}{KA} \frac{1}{\mathcal{A}}\iint_{\mathcal{M}}|P|^2\,d\mathcal{A},
\end{equation}
where the integrand is evaluated at $z=0$ and $d\mathcal{A}$ is an infinitesimal area element. In the supplementary material \cite{SM_ar}, we describe an elementary boundary element method (BEM) that we use to solve the homogenised model in the case of a finite metasurface.  

The homogenised model can be formally derived starting from the multiple-scattering formulation by applying the method of multiple scales in an appropriate limit process. Here we suffice with a heuristic justification. First, we assume that the effective ambient field $p\ub{a}_n$ experienced by the $n$th resonator can be approximated by the macroscale field $P$, so that
\begin{equation}\label{micro to macro} 
p_n= g P(\br_n).
\end{equation}
This implies that $p_n$ and $\alpha_n$ are slowly varying over the lattice of neck positions. The impedance condition \eqref{imp} then follows by comparing the normal macroscale `flux' $\mathscr{A}\partial{P}/\partial{z}$ at $z=0$ over an area $\mathscr{A}$ with the corresponding microscale flux $-2\pi \alpha_n (\mathscr{A}/A)$, and using \eqref{alpha p}. The linear dimension of the area $\mathscr{A}$ is taken to be large relative to the spacing $A^{1/2}$ yet small relative to the long scale $L$. The validity of this flux balance hinges upon the assumed slow variation of $P(\br)$, and hence $p_n$ and $\alpha_n$, as well as the assumption that the spacing is subwavelength. Expression \eqref{diss hom} for the averaged dissipation readily follows from \eqref{diss def} upon using \eqref{micro to macro}. We shall see that the long-scale assumption can fail at near-resonance frequencies for sufficiently low loss, owing to the excitation of short-wavelength surface waves, whereby the homogenised model losses validity.

\section{Critical coupling}\label{sec:cc}
\subsection{Single resonator}\label{ssec:cc_single}
In the case of a single resonator, the cavity pressure is $p_1= g p\ub{a}(\br_1)$. From \eqref{diss def}, we find using \eqref{g def}---\eqref{damping factors} that the dissipation at resonance is
\begin{equation}\label{D 1}
\bar{\mathcal{D}}_1=|p\ub{a}(\br_1)|^2\frac{2\pi}{\epsilon \bar\Omega^2 A}\frac{\bar\Delta_r\bar\Delta_v}{(\bar\Delta_r+\bar\Delta_v)^2}.
\end{equation}
For a fixed geometry, dissipation at resonance is optimised by tuning the loss parameter $\delta$ such that  $\bar{\Delta_v}=\bar{\Delta_r}$. We shall refer to this condition, which corresponds to a balance between dissipation and radiation damping as critical coupling for a single resonator. The corresponding explicit relation obtained in \cite{Brandao:20} is recovered by consistently approximating $\bar{\Omega}\sim K^{1/2}$.
From \eqref{damping factors}, critically coupling a single resonator demands very low loss, $\delta\simeq \epsilon^{5/2}$, resulting in a narrow resonance interval, $\Delta\Omega\simeq \epsilon^{3/2}$. 

\subsection{Infinite metasurface}\label{ssec:cc_inf}
Consider next the scattering problem where an infinite metasurface is exposed to a normally incident plane wave. Strictly speaking, scattering problems involving infinite metasurfaces can be ill-posed in frequency intervals where surface waves are supported, that is without generalising the radiation condition to account for this. For a dissipative metasurface, it seems sufficient to demand that the solution remains bounded. Under this assumption, symmetry of the metasurface and incident field implies that the cavity pressures in the multiple-scattering model are all equal, while the macroscale pressure in the homogenised model is a function of $z$ alone.  

Let us first assume the homogenised model. In light of the above, the macroscale pressure can be written 
\begin{equation}\label{P homo ref}
P=e^{-i\epsilon^{1/2}\Omega z}+\mathscr{R}e^{i\epsilon^{1/2}\Omega z},
\end{equation}
where the first term is the incident plane wave and the second term is a reflected plane wave, $\mathscr{R}$ being a reflection coefficient. The reflected plane wave should not be confused with the reflected field $p\ub{r}$, which in this case is the reflected wave with $\mathscr{R}=1$. The deviation of $\mathscr{R}$ from unity represents the aggregated effect of the spherical waves emitted from the necks. Indeed, the effective impedance condition \eqref{imp} gives the reflection coefficient $\mathscr{R}$ as 
\begin{equation}\label{R inf}
\mathscr{R}=\frac{A+ig\epsilon^{1/2}\Omega}{A-ig\epsilon^{1/2}\Omega}.
\end{equation}
Defining the collective radiation-damping factor
\begin{equation}\label{coll Del r}
\Delta_R=\frac{\epsilon^{1/2}\Omega K}{A},
\end{equation}
and noting that $\Delta_r\ll \Delta_R$, the reflection coefficient \eqref{R inf} can be approximated as 
\begin{equation}\label{R inf approx}
\mathscr{R}=\frac{\Omega^2-\bar{\Omega}^2-i\left(\Delta_R-\Delta_v\right)}{\Omega^2-\bar{\Omega}^2+i\left(\Delta_R+\Delta_v\right)}.
\end{equation}

In \cite{Brandao:20}, we derive an expression for $\mathscr{R}$ which is asymptotically equivalent to \eqref{R inf approx}, assuming only that $A\simeq 1$ as $\epsilon\to0$, i.e., that the spacing is subwavelength, specifically comparable to the characteristic cavity size. (Minor typos in \cite{Brandao:20} are corrected in the supplementary material \cite{SM_ar}.) Rather than relying on a homogenised model, the equivalent result in \cite{Brandao:20} is derived directly from the multiple-scattering model. The derivation entails the asymptotic approximation of a lattice sum. Comparison of these two distinct derivations confirms that, in the present special scenario of a plane wave normally incident on an infinite metasurface, the homogenisation approximation is valid for subwavelength metasurfaces regardless of the level of loss. We will see that this is not the case if either the metasurface is finite or the incident field is not a plane wave. 

According to \eqref{R inf approx}, $\mathcal{R}$ can be made to vanish at resonance by tuning the loss parameter $\delta$ such that $\bar{\Delta}_v=\bar{\Delta}_R$, where $\bar{\Delta}_R$ is $\Delta_R$ evaluated at $\Omega=\bar{\Omega}$. This is the condition for critical coupling of an infinite metasurface exposed to a normally incident plane wave. The corresponding explicit condition obtained in \cite{Brandao:20} is recovered by consistently approximating $\bar\Omega\sim K^{1/2}$. For a different perspective, consider the dissipation per resonator of the infinite metasurface, denoted $\mathcal{D}'_{\infty}$, which can be calculated in the homogenisation approximation using \eqref{diss hom} and \eqref{R inf}.  At resonance, $\mathcal{D}'_{\infty}$ is well approximated by
\begin{equation}
\bar{\mathcal{D}}'_{\infty}=\frac{4\bar{\Delta}_R\bar{\Delta}_v}{\left(\bar{\Delta}_R+\bar{\Delta}_v\right)^2},
\end{equation}
which can be optimised for a fixed geometry by tuning $\delta$ to satisfy $\bar{\Delta}_v=\bar{\Delta}_R$ at resonance, giving $\bar{\mathcal{D}}'_{\infty}=1$ as expected for perfect absorption and given our normalisation of the dissipation. As in the single-resonator case, the present critical coupling condition also corresponds to a balance between radiative and dissipative damping. The radiative damping in the metasurface case, however, is enhanced by an order  $1/\epsilon$ factor relative to the single-resonator case. Accordingly, critically coupling an infinite metasurface requires more loss, $\delta\simeq\epsilon^{3/2}$, resulting in a wider resonance, $\Delta \Omega\simeq \epsilon^{1/2}$. We also see from \eqref{R inf approx} that in the present idealised scenario excessive loss $\Delta_v\gg\Delta_R$ gives an effectively rigid substrate at resonance ($\mathscr{R}=1$), whereas deficient loss $\Delta_v\ll\Delta_R$ results in an effectively soft substrate at resonance ($\mathscr{R}=-1$).

\section{Surface waves}\label{sec:sw}
We continue to consider the case of an infinite metasurface, now looking for homogeneous surface-wave solutions that exist for real frequencies in the absence of an incident field and decay exponentially away from the substrate. Such solutions satisfy quasi-periodicity, i.e., that $p \exp \{-i\boldsymbol{\kappa}\bcdot\br\}$ is periodic over the two-dimensional lattice whose vertices are the limiting neck positions $\br_n$, wherein $\boldsymbol{\kappa}=\kappa_x\be_x+\kappa_y\be_y$ is a Bloch wave-vector parallel to the substrate plane with components $\kappa_x$ and $\kappa_y$ in the directions $\be_x$ and $\be_y$ parallel to the lattice vectors. For a prescribed real frequency $\Omega$, the Bloch wavevector $\boldsymbol{\kappa}$ is an eigenvector which for $\delta>0$ is expected to be complex valued, meaning that the surface waves exponentially attenuate parallel to the substrate plane. While defined for an infinite metasurface, we shall refer to these homogeneous solutions in the next section when interpreting the absorption characteristics of large metasurfaces. 

We first assume the homogenised model. In that case, the dispersion relation is isotropic, meaning that we can write $\boldsymbol{\kappa}=\kappa\unit$, $\kappa$ being a complex Bloch wavenumber and $\unit$ an arbitrary direction in the substrate plane. The reduced-wave equation \eqref{homo eq}, quasi-periodicity and the condition that the solution decays away from the substrate together imply that the macroscale pressure possesses the form 
\begin{equation}\label{macroscale Bloch decay}
P \propto \exp\left\{-\left(\kappa^2-\epsilon\Omega^2\right)^{1/2}z+i\boldsymbol{\kappa}\bcdot \br\right\},
\end{equation}
with the decay condition requiring $\ell_b^{-1}=\mathrm{Re}(\kappa^2-\epsilon\Omega^2)^{1/2}>0$, $\ell_b$ being the length scale of decay normal to the substrate. We also define the propagation length $\ell_p=1/\mathrm{Im}(\kappa)$, which is the length scale on which the surface wave attenuates parallel to the substrate plane. 
Given \eqref{macroscale Bloch decay}, the impedance boundary condition \eqref{imp} yields the dispersion relation $(\kappa^2-\epsilon\Omega^2)^{1/2}=\epsilon \Omega^2 g/A$. With this dispersion relation, the decay constraint and \eqref{g def} imply $\Omega<\bar\Omega$. 

Our use of the homogenised model is consistent only as long as it predicts surface wavelengths long compared to the spacing. To simplify the discussion below we assume $A\simeq1$, whereby the latter condition is $\kappa \ll 1$. The dispersion relation then gives $g\ll 1/\epsilon$, and so $\Delta_v\gg \epsilon$, i.e., $\delta\gg \epsilon^2$. In that regime, $\Delta_r\ll\Delta_v$ and the dispersion relation can be written 
\begin{equation}\label{homo sw}
\Omega^2-\bar{\Omega}^2+i\Delta_v+\frac{\epsilon K\Omega^2}{A(\kappa^2-\epsilon\Omega^2)^{1/2}}=0,
\end{equation}
with the shape factor $\sigma$ negligible in the expression \eqref{bar Omega} for $\bar\Omega$.  
\begin{figure}[t!]
\begin{center}
\includegraphics[trim = 100 0 0 0, scale=0.345]{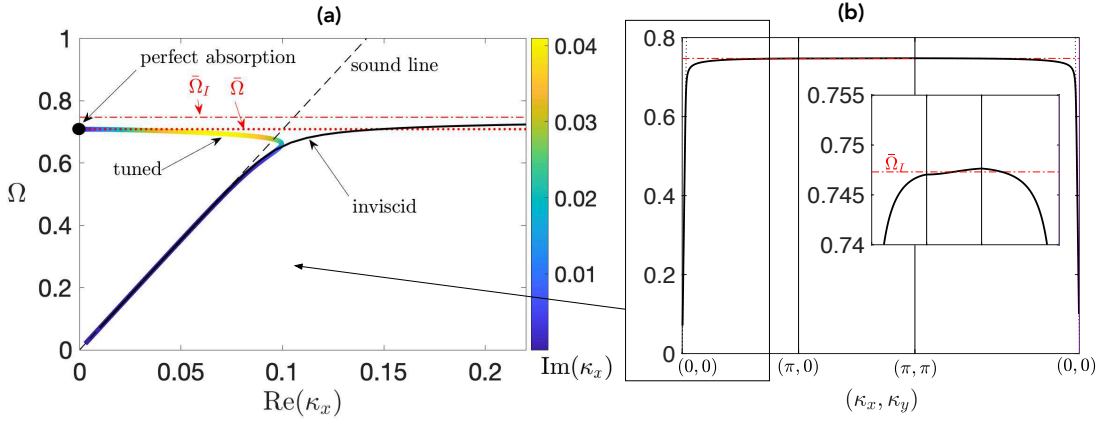}
\caption{Dispersion relation of surface waves supported by tuned (critically coupled) vs.~inviscid infinite metasurface. (a) Bloch wavevector in $x$ direction. Thick and coloured solid curve: tuned case, calculated from the complex-valued isotropic dispersion relation \eqref{homo sw} derived from the homogenised model. Thin solid curve: inviscid case, calculated from the real dispersion relation \eqref{sw inviscid} derived from the multiple-scattering model. (b) Inviscid dispersion relation continued around the edge of the irreducible Brillouin zone.}
\label{fig:SW}
\end{center}
\end{figure}

In particular, the homogenised model can be used to study surface waves under the infinite-metasurface critical-coupling condition, since then $\Delta_v\simeq \epsilon^{1/2}$. For that case, Fig.~\ref{fig:SW}a depicts the real and imaginary parts of $\kappa$ as a function of frequency, assuming for later comparison a wave in the $x$ direction: $\unit=\be_x$ and $\kappa=\kappa_x$. Consistently with the condition stated above for the homogenisation approximation to hold, the dispersion relation `folds' and $\kappa$ is small for all frequencies. As $\kappa\to0$, one branch joins with the sound line $\kappa=\epsilon^{1/2}\Omega$, whereas on the other branch $\Omega\to\bar{\Omega}$. In the limit $(\kappa,\Omega)\to(0,\bar{\Omega})$, we have $\ell_b\to\infty$ and the limiting state propagates to infinity, i.e., it is not a surface wave. In fact, this limiting state is nothing but the  solution already found in \S\ref{ssec:cc_inf} describing perfect absorption of a normally incident plane wave. 

Analysis of \eqref{homo sw} in the critically coupled case  $\Delta_v\simeq\epsilon^{1/2}$ reveals two distinguished regimes. The first, defined by the scalings $\kappa-\epsilon^{1/2}\bar{\Omega}\simeq\epsilon^{5/6}$ and $\Omega-\bar{\Omega}\simeq \epsilon^{1/3}$, corresponds to the dispersion curve separating from the sound line as the resonance frequency is approached. In that regime, $\ell_p\simeq 1/\epsilon$ and $\ell_b\simeq 1/\epsilon^{2/3}$, thus the waves propagate over large distances compared to the free-space wavelength but also decay away from the substrate on a scale larger than the free-space wavelength. In the second regime, defined by the scalings $\kappa\simeq \epsilon^{1/2}$ and $\Omega-\bar{\Omega}\simeq \epsilon^{1/2}$, viscosity enters the dominant balance and turns the dispersion curve around. In that regime, $\ell_p,\ell_b\simeq1/\epsilon^{1/2}$, thus the surface waves propagate and decay on length scales comparable to the free-space wavelength. 

For low-loss metasurfaces such that $\Delta_v$ is at most of order $\epsilon$, i.e., $\Delta_v=\mathcal{O}(\epsilon)$, the homogenisation model fails as it predicts surface waves with $\boldsymbol{\kappa}\simeq 1$. Similarly, the homogenisation model is not valid in scattering problems, whether involving a finite or infinite metasurface, where such short-wavelength surface waves are locally excited. Instead, we must return to the multiple-scattering formulation, where the quasi-periodicity condition becomes $p_n\propto \exp\left\{i\boldsymbol{\kappa}\bcdot \br_n\right\}$. In particular, in the supplementary material \cite{SM_ar} we show that in the limit $\epsilon\to0$, with $A\simeq1$, the dispersion relation for an inviscid metasurface is well approximated by 
\begin{equation}\label{sw inviscid}
\Omega^2-\bar\Omega_I^2+\frac{\epsilon K\Omega^2}{A\left(\kappa^2-\epsilon\Omega^2\right)^{1/2}}+\frac{\epsilon K^2}{2\pi}\left[\mathcal{S}_1(\boldsymbol{\kappa})+\mathcal{S}_2(\boldsymbol{\kappa})-\frac{2\pi}{A}\frac{1}{\kappa}\right]=0,
\end{equation}
where $\bar\Omega_I^2=K+\epsilon K^2\sigma$ is the inviscid resonance frequency and $\mathcal{S}_1(\boldsymbol{\kappa})$ and $\mathcal{S}_2(\boldsymbol{\kappa})$ are the absolutely convergent lattice sums
\refstepcounter{equation}
$$
\label{Sums}
\mathcal{S}_1(\boldsymbol{\kappa})=\sum_{\br\in \Lambda}'\frac{1+r}{r}e^{-r}\cos(\boldsymbol{\kappa}\bcdot \br), \quad 
\mathcal{S}_2(\boldsymbol{\kappa})=\frac{2\pi}{A}\sum_{\br\in\Lambda_*}\left(\frac{1}{|\boldsymbol{\kappa}-\br|}-\frac{2+|\boldsymbol{\kappa}-\br|^2}{(1+|\boldsymbol{\kappa}-\br|^2)^{3/2}}\right),
\eqno{(\theequation{\mathrm{a},\!\mathrm{b}})}
$$
in which $\Lambda=\{(n_1,n_2)\in\mathbb{Z}\, |\, A^{1/2}(n_1\be_x+n_2\be_y)\}$ is the infinite square lattice of neck positions, $\Lambda_*=\{(n_1,n_2)\in\mathbb{Z} \, | \, (2\pi/A^{1/2})(n_1\be_x+n_2\be_y)\}$ is the corresponding reciprocal lattice and the dash in (\ref{Sums}a) says to omit the zeroth lattice vector. In contrast to the critical-coupling regime, the dispersion relation in the inviscid case is real valued and anisotropic. In Fig.~\ref{fig:SW}, it is plotted for $\boldsymbol{\kappa}$ traversing the boundary of the reduced Brillouin zone \cite{Kittel:Book}. We see that the inviscid and critical-coupling cases, with all other parameters equal, agree for small $\kappa$ and close to the sound line. In the inviscid case, however, the dispersion curve does not fold but rather reaches $\kappa\simeq1$, where the dispersion curve is extremely flat, indicating small group velocity. In that flat-band regime, the surface wavelength is comparable to the subwavelength periodicity of the metasurface; the decay of the surface wave away from the substrate is also on that length scale, i.e., $\ell_b\simeq 1$; and, from \eqref{sw inviscid}, the dispersion relation can be approximated  explicitly: 
\begin{equation}
{\kappa}\simeq 1: \quad \Omega^2 -\bar{\Omega_I}^2 \sim -\epsilon\frac{K^2}{2\pi}\left(\mathcal{S}_1(\boldsymbol{\kappa})+{\mathcal{S}}_2(\boldsymbol{\kappa})\right).
\end{equation}
In the inviscid case the propagation length $\ell_p$ is  infinite. Analysing the propagation length of low-loss metasurfaces, $\Delta_v=\mathcal{O}(\epsilon)$, is challenging as it entails deriving a complex dispersion relation starting from the multiple-scattering theory. It is clear from the above, however, that in that case $1/\ell_p=\mathcal{O}(\epsilon)$.  
\begin{figure}[p!]
\begin{center}
\includegraphics[scale=0.195]{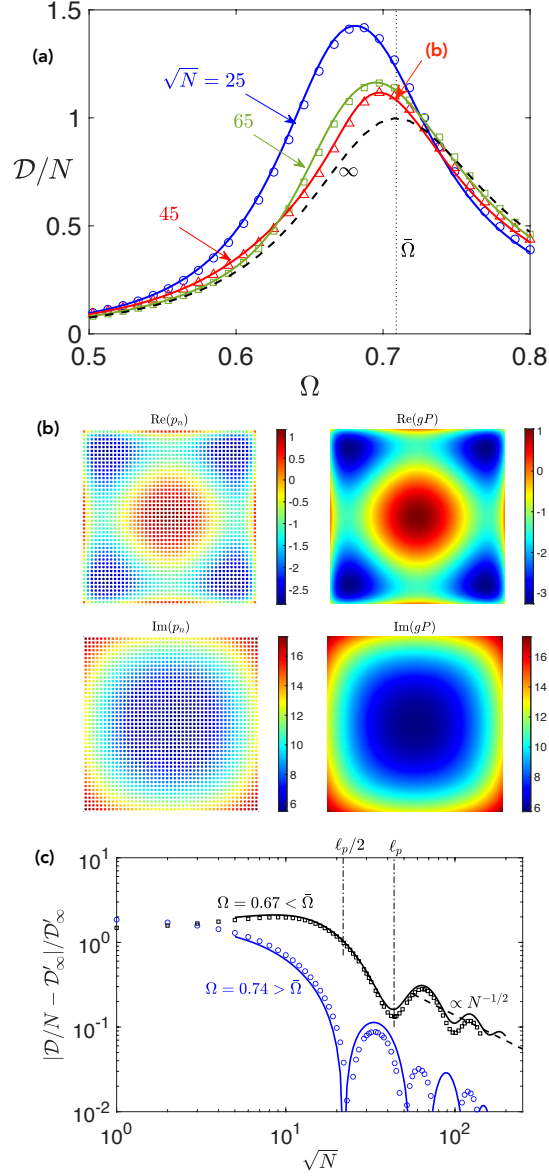}
\caption{Plane wave normally incident on finite metasurface with $\delta$ tuned to the critical-coupling condition for an infinite metasurface. (a) Averaged scaled dissipation per resonator, for several metasurface sizes. Symbols: multiple-scattering model. Solid curves: BEM solutions of homogenised model. Dashed curve: homogenised model for an infinite metasurface. (b) Real and imaginary parts of cavity pressures $p_n$ calculated from multiple-scattering model, alongside their continuous approximation from BEM solutions of the homogenised model, for $\sqrt{N}=45$ and $\Omega=\bar{\Omega}$. (c) Relative deviation of dissipation per resonator from theoretical value for infinite metasurface, as a function of metasurface size and for the indicated frequencies. Symbols: multiple-scattering model. Solid curves: BEM solutions of homogenised model.} 
\label{fig:cc_compare}
\end{center}
\end{figure}

\section{Large metasurfaces}\label{sec:large}
 \subsection{Tuned metasurface, plane-wave forcing}\label{ssec:large_tuned}
In this section, we present results for finite metasurfaces based on numerical solutions of the multiple-scattering model, and also of the homogenised model where relevant. We first consider the scenario where a plane wave is normally incident on a finite `tuned' metasurface. By tuned we mean that the loss parameter $\delta$ is tuned to the critical-coupling condition derived in \S\ref{ssec:cc_inf} assuming a plane wave normally incident on an infinite metasurface. 

Fig.~\ref{fig:cc_compare}a shows the absorption efficiency $\mathcal{D}/N$ as a function of the dimensionless frequency $\Omega$, for the indicated values of $N$. Values calculated based on the multiple-scattering model are seen to be in good agreement with BEM solutions of the homogenised model. Also shown is the theoretical approximation for an infinite metasurface, calculated using \eqref{diss hom}, \eqref{P homo ref} and \eqref{R inf}. By design, the efficiency of the infinite metasurface is unity at resonance, corresponding to perfect absorption. The finite metasurfaces are more efficient but, clearly, limited by their size. Qualitatively, their efficiency profiles approach that for an infinite metasurface slowly, non-uniformly (slower below the resonance frequency $\bar{\Omega}$) and non-monotonically. Furthermore, in contrast to the infinite-metasurface case, for large metasurfaces the cavity-pressure distribution can be appreciably non-uniform. Fig.~\ref{fig:cc_compare}b shows distributions calculated using the multiple-scattering model for $\sqrt{N}=45$, at resonance, alongside distributions obtained from BEM solutions of the homogenised model and using \eqref{coll Del r} to define continuous cavity-pressure distributions based on the macroscale pressure field.

Consider now the convergence with increasing metasurface size $\sqrt{N}$ of the efficiency of finite tuned metasurface to their infinite counterparts. Fig.~\ref{fig:cc_compare}f shows the relative difference in efficiency between the finite and infinite cases as a function of $\sqrt{N}$, for two values of $\Omega$. The lower frequency is $\Omega=0.67$. We have seen in \S\ref{sec:sw} that around this frequency an infinite tuned metasurface supports weakly bound, long-wavelength and attenuating surface waves, with the inverse surface-wavelength $2\pi/\kappa$, normal decay length $\ell_b$ and propagation length $\ell_p$ all comparable to the order $1/\sqrt{\epsilon}$ free-space wavelength. In our infinite-metasurface theory we have not included these surface-wave solutions as they are not bounded. For a large metasurface, these surface waves are excited, approximately, at the periphery and propagate inwards. This picture is consistent with the observation that the relative difference in efficiency remains order unity up to $\sqrt{N}\approx 20$, roughly $\ell_p/2$ for surface waves at that frequency. Between that size and  $\sqrt{N}\approx40$, roughly $\ell_p$, the relative difference decays by an order of magnitude, presumably owing to the the surface waves affecting an increasingly smaller proportion of the metasurface. For larger metasurfaces we see a much slower convergence, roughly like $1/\sqrt{N}$. The latter scaling is consistent with the effect of surface-wave excitation being limited to a neighbourhood of the periphery. At the higher frequency, $\Omega=0.74$, where an infinite tuned metasurface does not support surface waves, we observe a generally smaller relative difference in efficiency. 
\begin{figure}[p!]
\begin{center}
\includegraphics[trim = 60 0 0 0 0, scale=0.22]{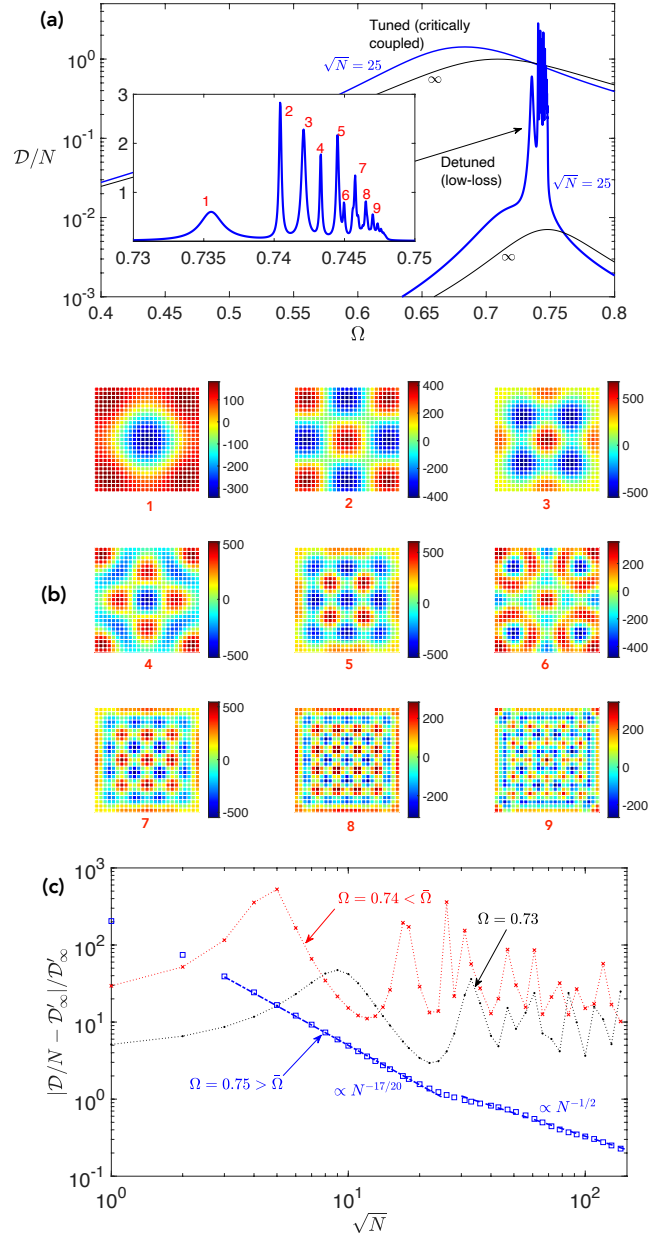}
\caption{Plane wave normally incident on tuned ($\delta$ tuned to critical coupling for an infinite metasurface) and detuned ($\delta$ tuned to critical coupling for a single  infinite metasurface) metasurfaces. (a) Averaged scaled dissipation per resonator for a finite metasurface of size $\sqrt{N}=25$, calculated using multiple-scattering model,  vs.~ homogenised model for an infinite metasurface. (b) Imaginary parts of the cavity pressures $p_n$ in the detuned case, at the meta-resonance peaks labeled in the inset of (a). (c) Relative deviation of dissipation per resonator from theoretical value for infinite detuned metasurface, as a function of metasurface size and for the indicated frequencies.}
\label{fig:cc_vs_detuned}
\end{center}
\end{figure}

\subsection{Detuned (low-loss) metasurface, plane-wave forcing}\label{ssec:large_detuned}
We next consider the scenario where a plane wave is normally incident on a finite `detuned' metasurface. By detuned we mean that the loss parameter $\delta$ is not tuned to the critical coupling condition for an infinite system. In particular, we shall specifically consider low-loss detuned metasurfaces for which $\delta$ is tuned to the single-resonator critical-coupling condition derived in \S\ref{ssec:cc_single}. According to the infinite-metasurface theory, such low-loss metasurfaces are extremely poor absorbers. In that theory, however, surface waves are excluded. In the detuned case, excitation of surface waves at the periphery can give rise to pronounced finite-size effects even for very large metasurfaces. Based on the inviscid multiple-scattering calculations in \S\ref{sec:sw}, for $|\Omega-\bar{\Omega}|\gg\epsilon$, with $\Omega<\bar{\Omega}$, we expect weakly bound surface waves similar to those in the tuned case. For $|\Omega-\bar{\Omega}|=\mathcal{O}(\epsilon)$, we expect surface waves that are strongly bound, short wavelength and hardly attenuate, with surface wavelength $2\pi/\kappa \simeq 1$, normal decay length $\ell_b\simeq 1$ and propagation length $\ell_p\gg1/\epsilon$. Note that the homogenisation model is not valid when such short-wavelength surface waves are excited. In this regime, surface waves can traverse even a large metasurface multiple times, reflecting and refracting at the periphery. For certain combinations of frequency and metasurface size, the surface waves may form standing-wave patterns over the square metasurface domain, similar to cavity modes of bulk waves. These constitute leaky states that are damped only by refraction at the periphery and the weak material loss.  

Fig.~\ref{fig:cc_vs_detuned}a shows the averaged dissipation per resonator, or absorption efficiency, plotted as a function of frequency, for tuned and detuned metasurfaces of finite size $\sqrt{N}=25$ and their theoretical infinite counterparts. The results for the finite metasurfaces are obtained by numerically solving the multiple-scattering model, while those for the infinite metasurfaces are calculated using \eqref{diss hom}, \eqref{P homo ref} and \eqref{R inf}. At most frequencies, the absorption efficiency of the detuned metasurfaces is smaller by orders of magnitude relative to the tuned metasurfaces, as one would expect from the infinite-metasurface theory. There is a narrow frequency interval below the resonance frequency $\bar{\Omega}$, however, where the detuned metasurfaces exhibit enhanced absorption, including a sequence of sharp peaks where efficiency is comparable and for some peaks even higher than that of the tuned metasurfaces.  We attribute these sharp peaks to collective meta-resonances of the metasurface as a whole, corresponding to excitation of the standing-wave leaky states anticipated above. This picture is qualitatively confirmed by Fig.~\ref{fig:cc_vs_detuned}b, which  shows cavity-pressure distributions computed at the first nine spectral peaks. These show standing-wave patterns of shorter and shorter surface wavelength as the cut-off frequency is approached, in accordance with the surface-wave dispersion curve shown in Fig.~\ref{fig:SW}b for an inviscid metasurface. 

For the detuned metasurfaces, Fig.~\ref{fig:cc_vs_detuned}c shows the relative difference in absorption efficiency between the finite and infinite cases as a function of $\sqrt{N}$, for three values of $\Omega$. The frequencies $\Omega=0.73$ and $\Omega=0.74$ are slightly  below the resonance frequency $\bar{\Omega}$ and the nearby surface-wave cut-off frequency. In contrast to the tuned case, we see no sign of convergence of the efficiency of finite detuned metasurfaces with increasing $N$, at least up to the values of $N$ we can access numerically. The relative difference remains very large (also because of the very small denominator) and oscillates wildly, more so at $\Omega=0.74$. Clearly, the apparent lack of convergence is a result of the long propagation length of surface waves at these frequencies, while the wild oscillations are associated with the excitation of meta-resonances, here accessed at fixed frequency by increasing size. The third frequency is $\Omega=0.75$; it is slightly above $\bar{\Omega}$ and the nearby surface-wave cut-off frequency. At that frequency, we observe monotonic convergence, the relative difference in efficiency vanishing like $1/N^{17/20}$ and, after $\sqrt{N}\approx 20$, like $1/N^{1/2}$. The latter scaling is consistent with local diffraction along the periphery   (not triggering surface waves). The origin of the former scaling is not clear. 
\begin{figure}[t!]
\begin{center}
\includegraphics[scale=0.4]{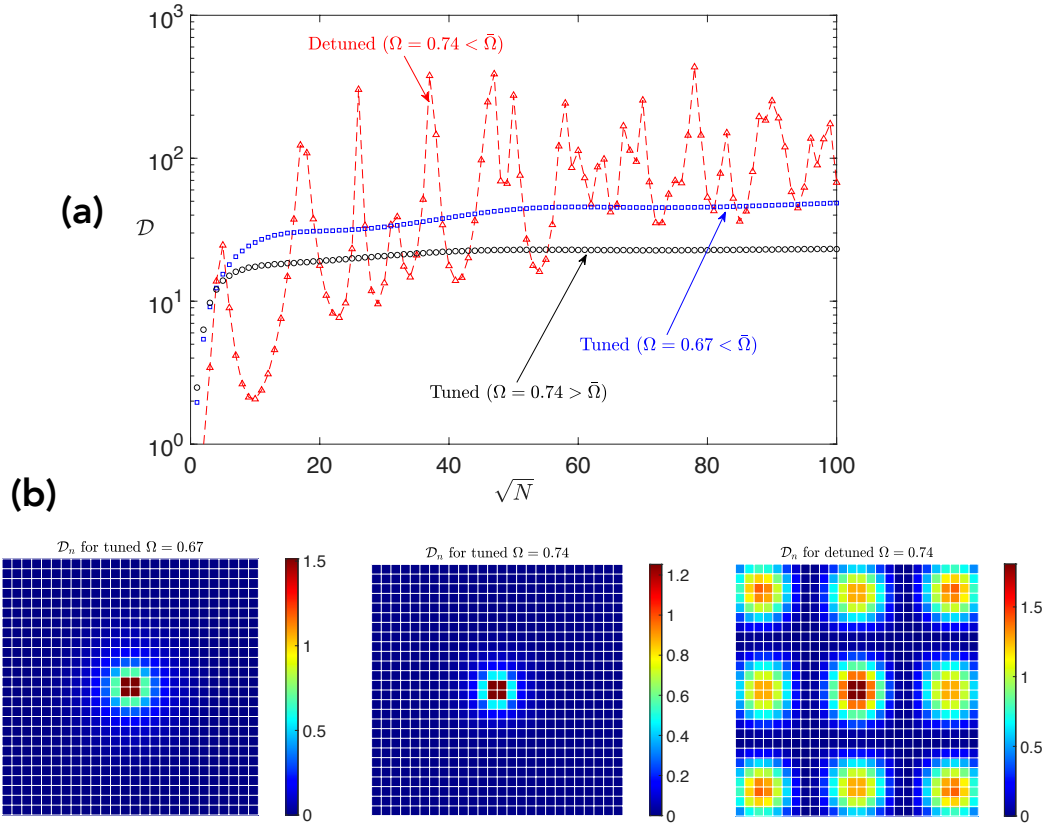}
\caption{(a) Total scaled dissipation for monopole source at unit height above the centre of tuned and detuned metasurfaces, as a function of metasurface size $\sqrt{N}$ and for the indicated frequencies. (b) Corresponding distributions of scaled dissipation $\mathcal{D}_n$, for $\sqrt{N}=26$.}
\label{fig:source}
\end{center}
\end{figure}
\subsection{Tuned and detuned metasurface, monopole-source forcing}
Consider now the scenario where instead of a normally incident plane wave there is a localised monopole source positioned at unit height above the centre of the metasurface. Here surface waves can be excited by the ambient field associated with the source acting on the bulk of the metasurface, as well as by diffraction of that ambient field at the periphery of the metsurface. Thus, in this  scenario surface waves are excited also for an infinite metasurface. We note that, even when short-wavelength surface waves are not excited, the homogenised model fails close to the source owing to its height above the substrate being comparable to the spacing. 

Fig.~\ref{fig:source}a shows the total dissipation as a function of $\sqrt{N}$, for tuned metasurfaces at $\Omega=0.67$ (below cut-off) and $\Omega=0.74$ (above cut-off) and for a detuned metasurface at $\Omega=0.74$ (below cut-off). For large tuned metasurfaces, dissipation at the lower frequency, where weakly bound and attenuating surface-waves are excited, is roughly $2.6$ higher than at the higher frequency, where surface waves are not excited. 
Large detuned metasurfaces, supposed to be poor absorbers, absorb more than large tuned metasurfaces, by an order of magnitude at meta-resonance peaks. Furthermore, the total dissipation oscillates wildly with increasing metasurface size. As in the plane-wave scenario, there is no indication of convergence up to the values of $N$ we compute. Fig.~\ref{fig:source}b contrasts the local dissipation near the source for a tuned metasurface with the distributed dissipation over a detuned metasurface excited at a meta-resonance.

\section{Concluding remarks}\label{sec:conc}
We have explored the effects of finite size on the absorption characteristics of large Helmholtz-type metasurfaces. The most significant effects are a consequence of the excitation of surface waves, and, for low enough material loss, meta-resonances associated with the surface waves forming standing-wave patterns over the metasurface domain. In particular, finite, even very large, low-loss metasurfaces exhibit absorption characteristics which are nothing like their theoretical infinite counterparts. Thus, while infinite low-loss metasurfaces are poor absorbers, large low-loss metasurfaces are efficient absorbers at near meta-resonance frequencies; this is in comparison with similar `tuned' metasurfaces satisfying the critical coupling condition for an infinite metasurface. The meta-resonances of low-loss metasurfaces are manifested in a series of sharp spectral peaks accumulating near the cut-off frequency for surface-wave propagation, which is close to the resonance frequency of an isolated resonator. This accumulation is understood to occur owing to the quantisation of short-wavelength surface waves, analogously to the accumulation of cavity modes at high frequencies. Thus, low-loss metasurfaces enable precise absorption of wave energy at a series of frequencies. This attribute may be exploited for filtering and sensing applications, assuming that the meta-resonance frequencies could be controlled at will.  

The present study has focused on modelling and numerical exploration. More sophisticated mathematical analysis is necessary in order to precisely predict the frequencies and quality factors of meta-resonances, as well as address several other open problems highlighted by our examples. These include studying the convergence, with increasing metasurface size, of absorption efficiency and other metasurface properties. As we have seen, the rate of convergence can be extremely non-uniform in frequency, and strikingly different for detuned (low-loss) and tuned (critically coupled) metasurfaces. A first step towards an improved understanding could involve asymptotic analysis of the homogenised problem. One relevant limit process is that of a metasurface comparable in size to the free-space wavelength, considered at a frequency and loss levels such that short-wavelength surface waves are excited. It may be easier to first consider a circular metasurface patch; asymptotic approximations could then be extracted either from an exact Wiener-Hopf solution, or, more insightfully, from a WKB approximation of the surface waves matched with local Wiener-Hopf solutions near the metasurface periphery \cite{Lam:89}, where the surface waves reflect and refract. Such an approach is only valid, however, in cases where the surface wavelengths are much larger than the periodicity of the metasurface; this must be kept in mind in order to avoid misleading predictions. Accordingly, a second step would be to address scenarios where periodicity-scale surface waves are excited. This could be done by a similar approach to that described above, with the significant technical complication of employing a two-scale WKB approximation \cite{Schnitzer:17} to describe the surface waves, matched with local discrete-Wiener-Hopf \cite{Tymis:14} solutions near the periphery. Of course, many generalisations are possible, including varying the shape of the metasurface domain, metasurfaces formed of non-identical resonators, interaction between multiple metasurface patches etc.

Lastly, since surface waves are intrinsic to metasurfaces formed of subwavelength resonators, we anticipate anomalous finite-size effects similar to what we have found for acoustic Helmholtz-type metasurfaces in other acoustic, as well as elastic and photonic, metasurface absorbers. To better understand these effects, our explicit, first-principles and computationally straightforward model could be employed in two ways. First, to design simple experiments to probe these finite-size effects in acoustics. Second, as a toy mathematical model for a generic metasurface absorber, ignoring the extensive  geometric and physical details encoded in the lumped parameters.

\bibliography{refs.bib}
\end{document}